\documentclass[12pt]{article}
\pdfoutput=1
\usepackage{graphicx, lutypaper,multirow}
\begin{document}

\newcommand{\ML}[1]{{\bf ML: #1}}


\newcommand\snowmass{\begin{center}\rule[-0.2in]{\hsize}{0.01in}\\\rule{\hsize}{0.01in}\\
\vskip 0.1in Submitted to the  Proceedings of the US Community Study\\ 
on the Future of Particle Physics (Snowmass 2021)\\ 
\rule{\hsize}{0.01in}\\\rule[+0.2in]{\hsize}{0.01in} \end{center}}

\begin{titlepage}

\snowmass

\title{Snowmass 2021 White Paper: Higgs Coupling Sensitivities and Model-Independent Bounds on the Scale of New Physics}

\author{Fayez Abu-Ajamieh}

\address{Center for High Energy Physics (CHEP), Indian Institute of Science (IISc)\\
C.V. Raman Avenue, Bangalore 560012 - India}

\author{Spencer Chang}

\address{Department of Physics and Institute for Fundamental Science\\ 
University of Oregon, Eugene, Oregon 97403}

\author{Miranda Chen, \ Da Liu, \ Markus A. Luty}

\address{Center for Quantum Mathematics and Physics (QMAP)\\
University of California, Davis, California 95616}

\begin{abstract}
In this Snowmass white paper, we describe how unitarity bounds can convert sensitivities for Higgs couplings at future colliders into sensitivities to the scale of new physics.  This gives a model-independent consequence of improving these sensitivities and illustrate the impact they would have on constraining new physics.  Drawing upon past successful applications of unitarity as a guide for future colliders (e.g.~the Higgs mass bound and discovering it at the LHC), we hope this data will be useful in the planning for next generation colliders.   
\end{abstract}
\end{titlepage}

The experimental study of the Higgs boson, 
in particular its couplings to Standard Model particles, is 
a crucial priority of the high-energy frontier.  
Unlike all other known elementary particles, the Higgs boson has no
quantum numbers that distinguish it from the vacuum, and plays
a fundamental role in the breaking of electroweak symmetry and
the origin of the mass of elementary particles.
Precise studies of the Higgs boson are critical for further
progress in elementary particle physics, and this is one
of the main motivations for next generation of high-energy
colliders.

In this Snowmass white paper, we focus on measurements of Higgs boson
couplings to Standard Model particles, specifically 
$W$, $Z$, $\gamma$, $g$, $t$, $b$, $\mu$, $\tau$, and the Higgs itself.
Comparisons of the sensitivity of future colliders for these
measurements is an important input in comparing different
proposals for future colliders.
For example,
Table 1 contains such a comparison, using numbers from the 
Higgs@FutureColliders study for the
European Strategy Report  \cite{deBlas:2019rxi}, as well as muon collider sensitivities from \cite{deBlas:2022aow, Forslund:2022xjq, DeBlas:2022wxr, Meade:2022}.

Within the Standard Model, all Higgs couplings are predicted at
high precision because they are related to other well-measured
quantities.
Therefore, in addition to being a crucial check of the Standard
Model, measurements of these couplings constitute a search for
physics beyond the Standard Model: 
any observed deviation from the Standard Model
prediction for these couplings is an unambiguous 
sign of new physics.
The most natural interpretation of such a deviation 
(assuming that no other new particles have been discovered)
is that the deviation is due to new particles and interactions
that are too heavy to be probed in current experiments.
In this scenario, perturbative unitarity is violated at high
energies.
This is because perturbative unitarity in the Standard
Model results from the cancelation of energy-growing
behavior in different amplitudes, and these cancelations
are spoiled when couplings deviate from their Standard Model
values.

For example, the right diagram of Fig.~1 shows a contribution to a $ZZ\to ZZZZ$
amplitude that involves the Higgs self-coupling $h^3$.
This diagram by itself violates perturbative unitarity at
high energies, but in the Standard Model
the leading high-energy behavior of this diagram
is canceled by additional diagrams such as those shown in the left diagram of 
Fig.~1, which do not depend on the Higgs cubic coupling.
For this reason, any deviation in the $h^3$ coupling compared
to the canceling diagrams leads
to violation of perturbative unitarity at high energies.
In \Refs{Chang:2019vez, Falkowski:2019tft, 
Abu-Ajamieh:2020yqi, Abu-Ajamieh:2021egq, Abu-Ajamieh:2022ppp} the leading high-energy behavior of these 
amplitudes were computed using equivalence
principle techniques, and unitarity bounds were presented
for various Higgs couplings.

\begin{figure*}[t]
\begin{center}
\begin{minipage}{3.75in}
\begin{center}
\includegraphics[width=\linewidth]{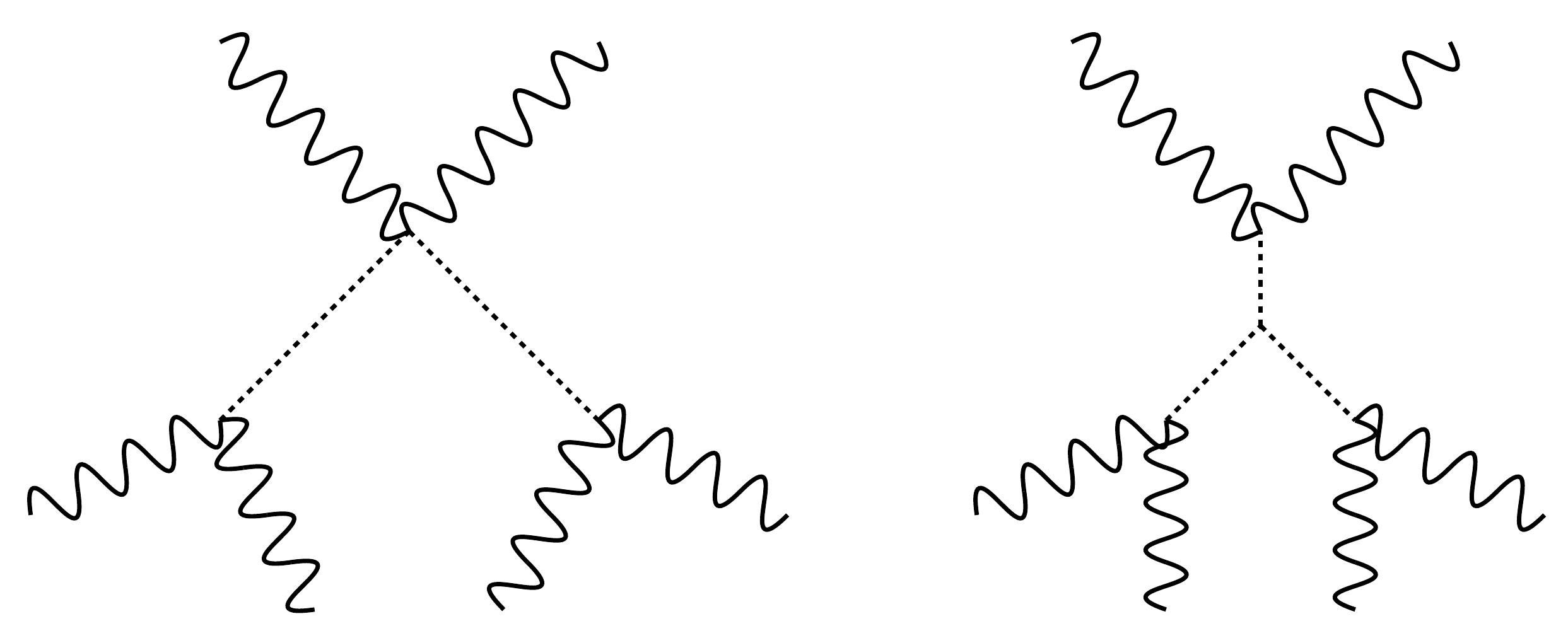}
\end{center}
\small\caption{Feynman diagrams for $ZZ\to ZZZZ$
in unitary gauge.
\label{fig:sixZs}}
\end{minipage}
\end{center}
\end{figure*}

In this way, the sensitivity of Higgs coupling
measurements can be directly 
translated to a a sensitivity to the scale $\La$ of new 
physics that can give rise to deviations from the Standard Model prediction.
We emphasize that this connection is completely model-independent,
since it only uses the measured values of couplings and the assumed
absence of new physics below $\La$ to determine the scale of unitarity
violation.
In particular, any observed deviation in the couplings gives an upper
bound on the scale of new physics.
In many cases, this is a scale that can be explored in
future collider experiments.

We present the bounds on the scale $\La$ 
of new physics arising from Higgs coupling
measurements at various future colliders in Table 1.  A graphical depiction of these numbers can be seen in Figure 2.
For the deviations considered here, the unitarity bounds from the $h$ coupling to $X$
scale as $\La_X \propto (\de\ka_X)^{-1/2}$, so increasing
sensitivity by a factor of 4 gives a factor of 2 improvement on the
bound of new physics.
The numerical values involve the scale where couplings become
strong, and is therefore subject to theoretical
uncertainties, which can be estimated by varying the unitarity bound on the amplitude (see \cite{Abu-Ajamieh:2020yqi, Abu-Ajamieh:2021egq, Abu-Ajamieh:2022ppp}).  
For example, varying the bound on the amplitude by a factor of 4 would change the unitarity bound by a factor of 2.  
Despite these uncertainties, they provide a critical model-independent estimate of
the scale of new physics that can motivate future experiments,
just as the unitarity bounds on the Higgs boson mass motivated
the design of the Large Hadron Collider, which ultimately discovered the
Higgs boson well below the unitarity bound.
To conclude, the model-independent bound on the scale of new physics probed by
these measurements gives a physical interpretation of the sensitivity
of these measurements that is complementary to the comparison with
specific models.

\begin{table}
\setlength{\tabcolsep}{1pt}
\renewcommand{\arraystretch}{1.3}
{\footnotesize
\begin{tabular}{| c | c | c | c  c | c cc | c c c | c |  c c |  c | c c |}
\hline
Coupling (2$\sigma$) &HL-LHC   &LHeC   &\multicolumn{2}{c|}{HE-LHC}  & \multicolumn{3}{c|}{ILC}   & \multicolumn{3}{c|}{CLIC}     &CEPC  & \multicolumn{2}{c|}{FCC-ee}  &FCC & \multicolumn{2}{c|}{Muon} \\
Unitarity Bound &   &   &S2 & S2$^\prime$   &250 & 500 & 1000 & 380 & 1500 & 3000 & & 240 & 365 & ee/eh/hh & 10 & 10 TeV+   \\ 
& & & & & & & & & & & & & & &TeV & 125 GeV \\
\hline
$2\delta \kappa_{V}$ [\%] & $3.0$   & 1.5 & $2.6$   & 1.8 & $0.58$   & $0.46$ & 0.44 & $1.0$   & $0.32$   & $0.22$   & $0.28$   & $0.40$   & $0.34$   & $0.24$ & $0.26$ & $0.24$ \\ 
$\Lambda_{V}$ (TeV) & 6.0 & 9 & 6.4 & 7.7 & 14 & 15 & 16 & 10 & 18 & 22 & 20 & 16 & 18 & 21& 20 & 21 \\ \hline
2$\delta\kappa_{g}$ [\%]  & $4.6$   & $7.2$   & $3.8$ & 2.4 & $4.6$   & $1.94$ & 1.32  & $5.0$   & $2.6$   & $1.8$   & $3.0$   & $3.4$   & $2.0$   & $0.98$  & $1.34$ & $1.31$ \\
$\Lambda_{g}$ (TeV) & 51 & 41 & 56 & 70 & 51 & 78 & 95 & 49 & 68 & 81 & 63 & 59 & 77 & 110 & 94 & 95 \\ \hline 
2$\delta\kappa_{\gamma}$ [\%]  & $3.8$   & $15.2$   & $3.2$ & 2.4  & $13.4$   & $6.8$ & 3.8  & $196$  & $10$   & $4.4$   & $7.4$   & $9.4$   & $7.8$   & $0.58$ & $2.2$ & $2.13$ \\
$\Lambda_{\gamma}$ (TeV) & 120 & 61 & 130 & 150 & 65 & 92 & 120 & 17 & 76 & 110 & 88 & 78 & 86 & 310 & 160 &160 \\ \hline
2$\delta\kappa_{Z\gamma}$  [\%]  & $20$   & $-$   & 11.4 & 7.6 & $198$   & $172$ & 170  & $240$   & $30$   & $13.8$   & $16.4$   & $162$   & $150$   & $1.38$  & $20$ & $20$ \\
$\Lambda_{Z\gamma}$ (TeV) & 34 & $-$ & 45 & 55 & 11 & 12 & 12 & 10 & 28 & 41 & 37 & 12 & 12 & 130& 34 & 34 \\ \hline

$2\delta \kappa_{t}$ [\%]  & $6.6$   & $-$   & 5.6 & 3.4 & $-$   & $13.8$ & 3.2  & $-$   & $-$   & $5.4$   & $-$   & $-$   & $-$   & $2.0$& $104$ & $4.2$ \\
$\Lambda_{t}$ (TeV) & 13 & $-$ & 14 & 18 & $-$ & 9 & 19 & $-$ & $-$ & 14 & $-$ & $-$ & $-$ & 24 & 3 &16 \\ \hline
$2\delta \kappa_{b}$  [\%]  & $7.2$   & $4.2$   & 6.4 & 4.6 & $3.6$   & $1.16$ & 0.96  & $3.8$   & $0.92$   & $0.74$   & $2.4$   & $2.6$   & $1.34$   & $0.86$ & $0.54$ & $0.48$ \\
$\Lambda_{b}$ (TeV) & 80 & 100 & 85 & 100 & 110 & 200 & 220 & 110 & 220 & 250 & 140 & 130 & 180 & 230& 290 & 310 \\ \hline
$2\delta \kappa_{\mu}$  [\%]  & $9.2$   & $-$   & 5.0 &  3.4 & $30$   & $18.8$ & 12.4  & $640$   & $26$   & $11.6$   & $17.8$   & $20$   & $17.8$   & $0.82$ & $3.6$ & $0.19$ \\
$\Lambda_{\mu}$ (TeV) & 590 & $-$ & 800 & 970 & 320 & 410 & 510 & 70 & 350 & 520 & 420 & 400 & 420 & 2000& 540 & 2400 \\ \hline
$2\delta \kappa_{\tau}$ [\%]   & $3.8$   & $6.6$   & $3.0$ & 2.2 & $3.8$   & $1.40$  & 1.14  & $6.0$   & $2.6$   & $1.76$   & $2.6$   & $2.8$   & $1.46$   & $0.88$ & 0.47& 0.47\\
$\Lambda_{\tau}$ (TeV) & 220 & 170 & 250 & 290 & 220 & 370 & 410 & 180 & 270 & 330 & 270 & 260 & 360 & 460& 360 & 360 \\ \hline
$2\delta \kappa_{h}$ [\%]   & $94$   & $-$   & $40$ & 40 & $58$    & 54  & 20 & 92   & 72   & 22   & 34   & 38   & 38   & 10 & 7.4 & 7.4 \\
$\Lambda_{h}$ (TeV) & 15 & $-$ & 23 & 23 & 19 & 19 & 32 & 15 & 17 & 30 & 25 & 23 & 23& 45 &  52 & 52 \\ \hline
\end{tabular}
}
\caption{Approximate 
$2\sigma$ sensitivities for Higgs couplings obtained by doubling the $1\si$
sensitivities reported in the
Higgs@FutureColliders study \cite{deBlas:2019rxi}.  The muon collider numbers use numbers from \cite{deBlas:2022aow,Forslund:2022xjq, DeBlas:2022wxr, Meade:2022}.  
The sensitivities are obtained under the assumption that there are no additional light
states that the Higgs decays into (`kappa-0 scenario').  
For $\de\kappa_V$ we use the smaller of $\de\kappa_W$ and $\de\kappa_Z$, while for 
$\de\kappa_h$ we use the best sensitivity reported for single or di-Higgs production.  
Below each sensitivity line, we list the unitarity bound for a Higgs coupling bound, 
$\Lambda_X$, assuming $\kappa_X = 1+2\delta \kappa_X$.   
Due to theoretical uncertainties on the unitarity bound, we present the bounds 
to two significant digits.    
\label{tbl:couplings}}
\end{table}

\begin{figure}[tbh]
\centering
\hspace{3cm} \includegraphics[width=0.95\linewidth]{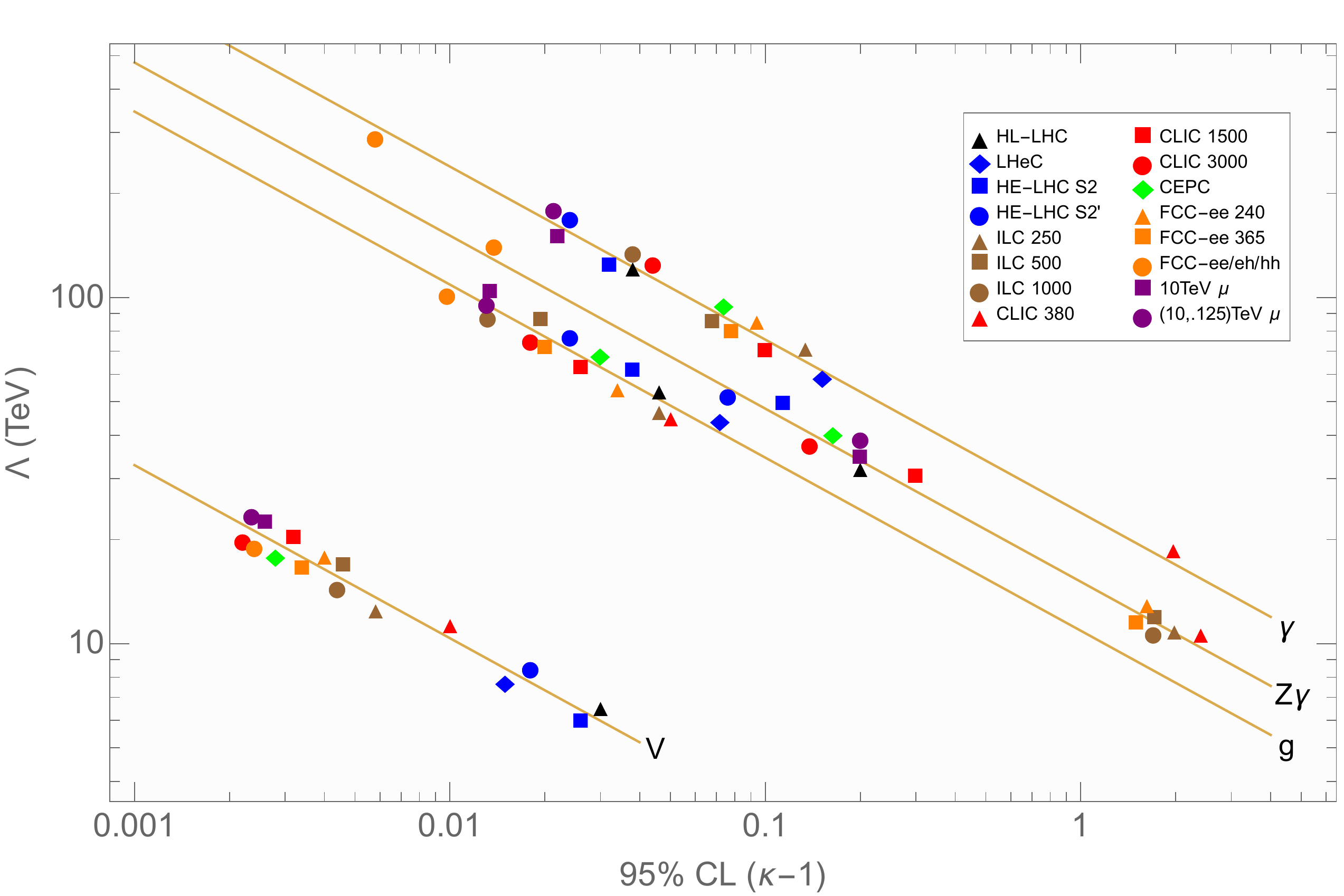}
\includegraphics[width=0.9505\linewidth]{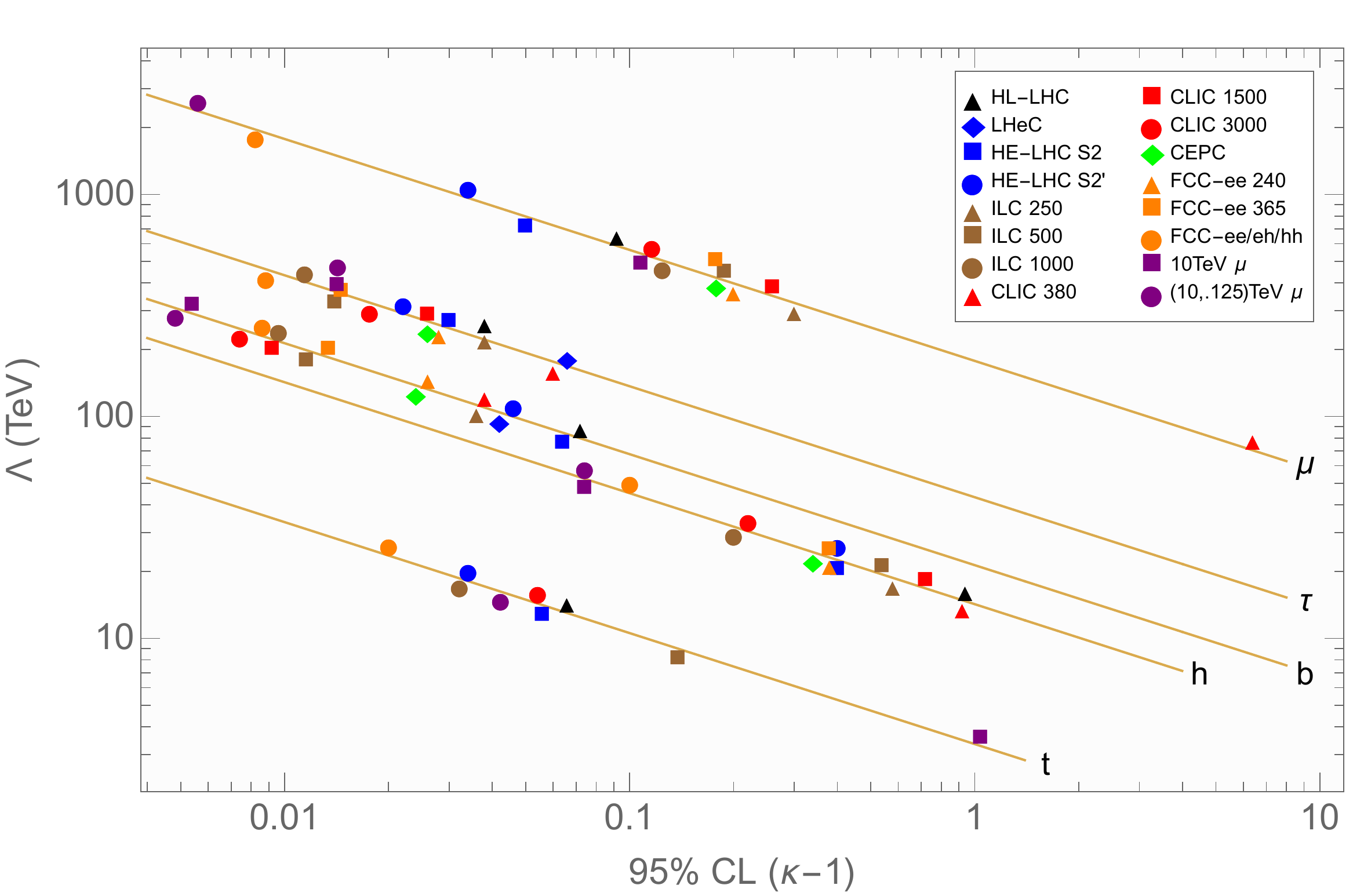}
\caption{\label{fig:plots} Figures for the data in Table 1.  The lines represent the theoretical relation between the precision on $\kappa - 1$ and the scale of new physics.   All data points should be on the line, but are arbitrarily displaced up and down for visual clarity.}
\end{figure}

\section*{Acknowledgements}
FA is supported
by the C.V. Raman fellowship from CHEP at IISc., SC is supported in
part by the U.S. Department of Energy under Grant Number DE-SC0011640, while ML, DL and MC are supported in part by 
the U.S. Department of Energy under grant DE-SC-0009999.

\bibliographystyle{utphys}
\bibliography{Kappa_Unitarity}

\end{document}